# Periodic transmission peak splitting in one dimensional disordered photonic structures

Ilka Kriegel[1], Francesco Scotognella[2,3]*

[1]Department of Nanochemistry, Istituto Italiano di Tecnologia (IIT), via Morego, 30, 16163 Genova, Italy

[2]Dipartimento di Fisica, Politecnico di Milano, Piazza Leonardo da Vinci 32, 20133 Milano, Italy

[3]Center for Nano Science and Technology@PoliMi, Istituto Italiano di Tecnologia, Via Giovanni Pascoli, 70/3, 20133, Milan, Italy

* Corresponding author at: Dipartimento di Fisica, Politecnico di Milano, Piazza Leonardo da Vinci 32, 20133 Milano, Italy

E-mail address: francesco.scotognella@polimi.it (F. Scotognella)

**Abstract**

In the present paper we present ways to modulate the periodic transmission peaks arising in disordered one dimensional photonic structures with hundreds of layers. Disordered structures in which the optical length *nd* (*n* is the refractive index and *d* the layer thickness) is the same for each layer show regular peaks in their transmission spectra. A proper variation of the optical length of the layers leads to a splitting of the transmission peaks. Notably, the variation of the occurrence of high and low refractive index layers, gives a tool to tune also the width of the peaks. These results are of highest interest for optical application, such as light filtering, where the manifold of parameters allows a precise design of the spectral transmission ranges.

**Introduction**

Besides periodic photonic crystals [1,2], disordered photonic structures have attracted increasing attention because of their intriguing optical properties [3–5]. Such disordered media have been exploited for random lasers [6–8] and light harvesters to improve solar cell efficiency [9,10]. The case in which the disorder occurs only in one dimension (1D), neglecting the propagation of light in the other two dimensions, is favourable in terms of simple fabrication and modelling. In 1D disordered photonic structures interesting phenomena as Anderson localization [11] and optical necklaces [13] have been observed. In long 1D disordered photonic structures (e.g. hundreds of layers), other peculiar effects have been predicted: i) the total transmission, as a function of the maximum size of the high refractive index clusters, shows a valley followed by oscillations [14]; the transmission spectra of random structures in which the layers have the same optical length ($n_1 \cdot d_1 = n_2 \cdot d_2$, with *n* being the respective refractive index and *d* the layer thickness) show periodic peaks [12].

In this paper we show that the periodic peaks arising in disordered 1D photonic structures can be engineered by a proper control of the layer thickness (i.e. the optical length *nd* of the layers). Slight increase of the optical length while keeping the refractive index constant leads to a splitting of the periodic peaks, and the splitting is proportional to the variation of the optical length of the layers. Furthermore, the ratio between high and low refractive index layers in the structure allows the control over the width of the transmission peaks. Together, these parameters enable a targeted design of the transmission properties of the disordered 1D photonic crystal. This is of highest significance to the optics community as the large variety of optical design parameters goes in hand with facile fabrication of the 1D photonic structure.

**Methods**

The one-dimensional disordered photonic structures have been realized with two materials with refractive indexes $n_1$=2.4 and $n_2$=1.6, respectively. In this work we did not consider materials with energy dependent refractive indexes. However, due to the low dispersion in the visible only minor changes to the optical properties are expected in the observed spectral range. The refractive index of each layer was kept constant throughout the calculations and only the layer thickness has been varied as: $d_1=(310/x_1)$ and $d_2=(310/x_2)$, where $x_1=n_1$, $n_1$-0.05, $n_1$-0.1,…2.2, and $x_1=n_2$, $n_2$+0.05, $n_2$+0.1,…1.8, i.e. from $d_1=(310/2.4)$nm and $d_2=(310/1.6)$nm, respectively, to $d_1=(310/2.2)$nm and $d_2=(310/1.8)$nm, respectively. Thus, the same optical thickness for both layers is only valid for $x_1=n_1$ and $x_2=n_2$. A deviation of the optical length is introduced for the other structures ($n_1d_1 \neq n_2d_2$), with increasing deviation the larger the variation of $x_1$ and $x_2$. The total number of layers for the photonic structures is 600. The disorder has been introduced by inducing a 50% probability for each layer to appear. For details the reader is referred to reference [12].

To simulate the transmission spectra of the one dimensional photonic structures we have employed the transfer matrix method [14–16], calculating the transmission values each 0.5 meV in the range 1.5 – 6.5 eV.

**Results and Discussion**

In Figure 1 we show the with the transfer matrix method calculated transmission spectra of 1D disordered photonic structures with different optical length ($nd$, where $n$ is the refractive index and $d$ the layer thickness). The refractive indexes of the two materials are $n_1$=2.4 and $n_1$=1.6 throughout all calculations. Close to each spectrum, the two numbers $x_1/x_2$ are related to the thickness of the layers through the relation $d_1=(310/x_1)$nm and $d_2=(310/x_2)$nm. In this way, the top spectrum refers to a structure with $d_1=(310/2.4)$nm and $d_2=(310/1.6)$nm, where the optical thickness of the two layers is the same, $n_1d_1=n_2d_2$. The other spectra show a slight deviation of the optical length, $n_1d_1 \neq n_2d_2$, as well explained in the Methods section above. For each of the 600 layers of the structure, the probability to have the material with $n_1$ or the material with $n_2$ is 50%.

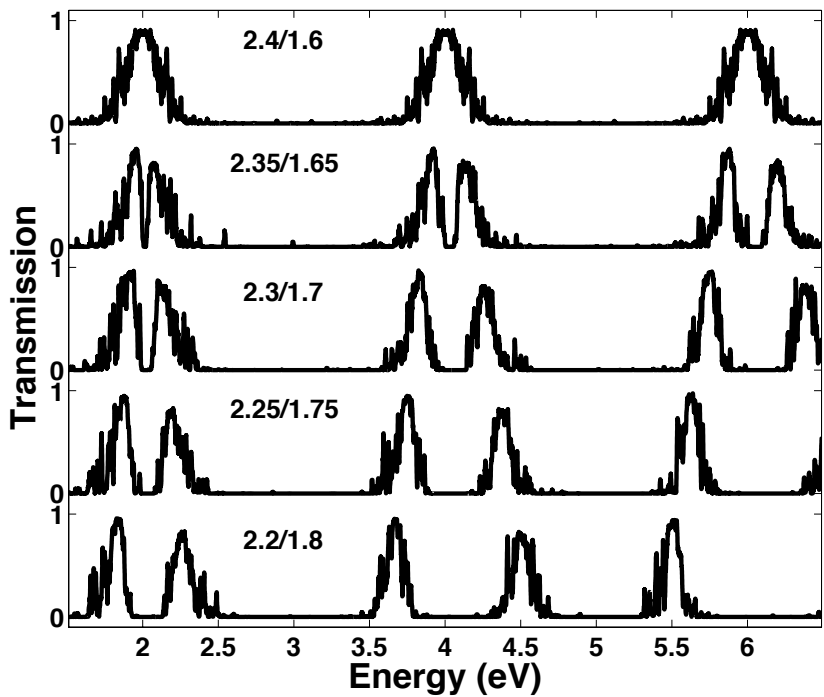

**Figure 1.** Transmission spectra of 600 layer disordered photonic structures with different layer thicknesses of the two materials inducing a deviation of the optical length of both layers. For each spectrum the layers with refractive index $n_1$=2.4 have thickness $d_1=(310/x_1)$nm while the layers with refractive index $n_2$=1.8 have thickness $d_2=(310/x_2)$nm. $x_1$ and $x_2$ are displayed close to the spectrum. For example, the first spectrum is related to a structure with $d_1$=(310/2.4)nm and $d_1$=(310/1.6)nm. Only for the first transmission spectrum the condition for the same optical thickness $n_1d_1=n_2d_2$ is fulfilled, while for the other $n_1d_1 \neq n_2d_2$, with increasing deviation of the optical length from top to bottom.

In the studied region, between 1.5 and 6.5 eV, intense transmission peaks are situated at 2, 4, and 6 eV for the spectrum with the same optical thickness (2.4/1.6). The observed periodicity of the transmission peaks has previously been studied by us (Ref. [12]). Thereafter, the layer thickness for both the materials has been varied with increasing deviation of their optical thickness from top to bottom, i.e. from 2.35/1.65 to 2.2/1.8. Remarkably, this deviation induces a dip in the transmission leading to a splitting of each transmission peak into two peaks. As shown in Figure 1, the intensity of the splitting is increasing with increasing deviation from of the layer thickness, i.e. the layer optical length.

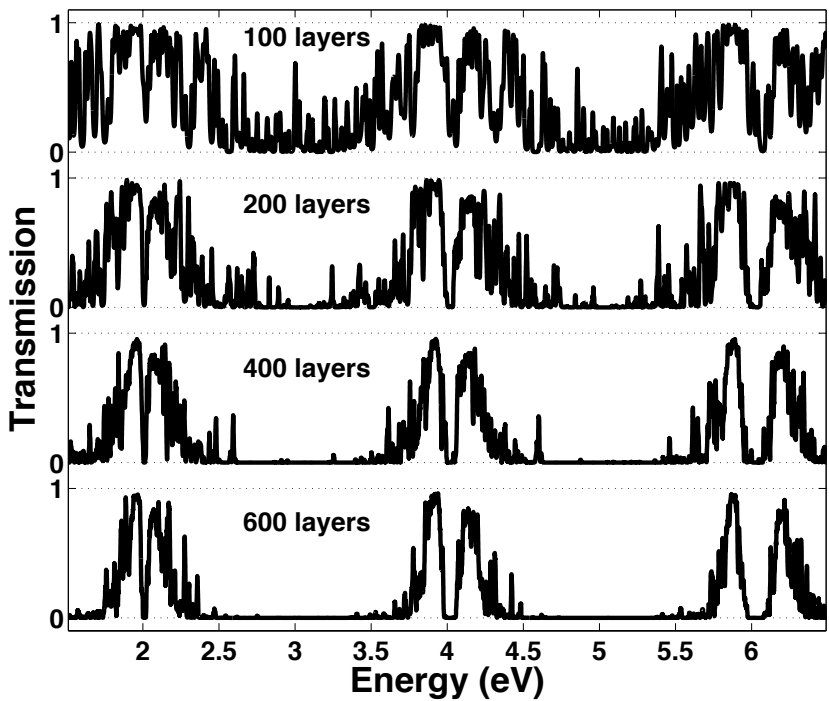

**Figure 2.** Transmission spectra of the disordered photonic structure, with $d_1$=(310/2.35)nm and $d_2$=(310/1.65)nm, for different numbers of layers.

Focussing on one of the spectra in Figure 1 [e.g. the one with $d_1$=(310/2.35)nm and $d_2$=(310/1.65)nm], we observe that the peak splitting is present also for a smaller number of layers. In Figure 2 it is illustrated that already for as low as 100 layers a transmission dip at 4 eV is noticeable, although, a higher number of layers results in more clearly defined split peaks, and a negligible light transmission in the other regions of the spectra.

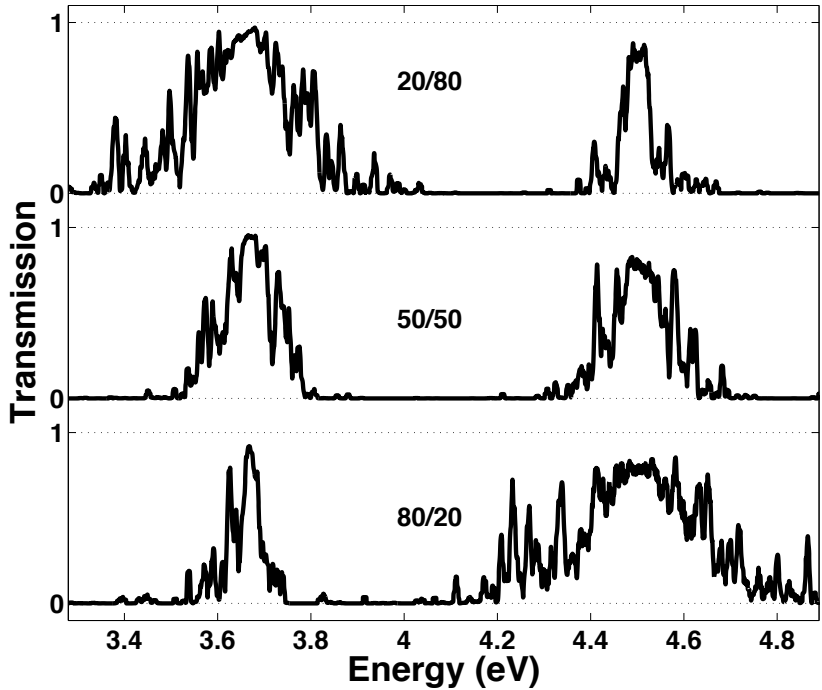

**Figure 3.** Transmission spectra, around 4 eV, of the 600 layer disordered photonic structure with $d_1$=(310/2.2)nm and $d_2$=(310/1.8)nm. The numbers for the three spectra represent the probability for the occurrence of the respective layer in the photonic structure, i.e. the material with $n_1$ (e.g. 20% for the top spectrum) or the material with $n_2$ (e.g. 80% for the top spectrum).

Notably, an additional means of control of the light transmission in these 1D structures is given by varying the probability for each of the two layer materials to occur in the 1D structure. When the two materials have the same probability (as for Figure 1 and 2) then the width of the two split peaks is identical (Figure 3, central spectrum). This can be influenced by changing the probability for each layer. The top spectrum in Figure 3 refers to the transmission of a structure in which the probability to have layers with high refractive index ($n_1$) is 20%, while the probability to have layers with low refractive index ($n_2$) is 80%. Focussing on the split peaks around 4 eV, the width of the low energy peak is broadened while the width of the high energy one is narrowed. Interestingly, the situation is inverted by inverting the probability for the occurrence of the two materials in the structure (Figure 3, spectrum at the bottom).

**Conclusions**
In this study we have shown the possibility to engineer the periodic transmission peaks arising in 1D disordered photonic structures where the layers have the same optical length. A proper variation of the optical length gives rise to a splitting of such periodic peaks, and the intensity of the splitting is directly related to the layer optical length with increasing intensity for an deviation from of the layer optical length. In addition, we have noticed that the occurrence of high and low refractive index layers in the structure influences the width of the split transmission peaks, inducing an asymmetry of the split peaks by changing the probability for their occurrence. To experimentally confirm the study, we suggest the fabrication of such 1D disordered photonic structures with rf sputtering [17,18], that ensures a high optical quality with a large number of layers. Such structures can be interesting for the realization of high precision optical filters for the selection of a specific optical signal with a precise spectral width.